\documentclass[twocolumn,aps,pra]{revtex4-1}
\usepackage{dcolumn}    
\usepackage[T1]{fontenc}
\usepackage[utf8]{inputenc}
\usepackage{geometry,array,graphicx}
\usepackage{geometry,amsmath}
\usepackage{amssymb} 
\usepackage{color,graphicx}
\usepackage{array}
\usepackage{bm}         
\usepackage{ifpdf}
\usepackage[english]{babel}
\usepackage{multirow}
\usepackage{epstopdf}
\usepackage{tabularx}
\usepackage{mathrsfs}
\usepackage{upgreek}
\usepackage{mathtools}
\usepackage{epstopdf}
\usepackage{hyperref}
\usepackage{color}

\newcommand{\ket}[1]{\vert #1 \rangle}
\newcommand{\bra}[1]{\langle #1 \vert}
\newcommand{\expv}[1]{\langle #1 \rangle}

\begin{document}

\title{A hybrid scheme for prime factorization and its experimental implementation using IBM quantum processor}

\author{Ashwin Saxena$^{\dagger,1}$, Abhishek Shukla$^{\dagger,\ddagger,2}$  and Anirban Pathak$^{\dagger,3}$}
\email{$^1$ ashn5.new@gmail.com}
\email{$^2$ abhishek.shukla@mail.huji.ac.il}
\email{$^3$ anirban.pathak@jiit.ac.in}

\affiliation{$^\dagger$ Jaypee Institute of Information Technology, A 10, Sector 62, Noida, UP 201309, India}
\affiliation{$\ddagger$ Department of Applied Physics, Rachel and Selim school of Engineering, The Hebrew University of Jerusalem, Jerusalem 91904, Israel}

\begin{abstract}
We report a quantum-classical hybrid scheme for factorization of bi-prime numbers (which are odd and square-free) using IBM's quantum processors. The hybrid scheme proposed here involves both classical optimization techniques and adiabatic quantum optimization techniques, and is build by extending a previous scheme of hybrid factorization [Pal et al., Pramana 92, 26 (2019) and Xu et al., Phys. Rev. Lett. 108, 130501 (2012)]. The quantum part of the scheme is very general in the sense that it can be implemented using any quantum computing architecture. Here, as an example, we experimentally implement our scheme for prime factorization using IBM's QX4 quantum processor and have factorised 35.
\end{abstract}

\maketitle

\section{Introduction} \label{intro}

It is well known that prime factorization is a computationally difficult problem and the security of the RSA-type classical cryptographic systems  derive from this difficulty \cite{rivest1978method}. Although, various RSA-type schemes for public key cryptography are still in use, the confidence in the
security provided by RSA-type cryptosystem has been considerably weakened since the pioneering work of Shor \cite{shor1999polynomial,shor1995scheme,shor2000simple,shor1994algorithms}. Specifically, in \cite{shor1999polynomial}, Shor showed that the factorization of bi-primes can be performed in polynomial time if a scalable quantum computer can be built. In other words, building of a scalable quantum computer would imply the end of RSA-type classical cryptosystem and a set of other classical cryptosystems, too. This fundamental importance of the factorisation problem and the benefit
of implementing it using a quantum computer have led to a set of schemes for prime factorization, mainly experimental \cite{Vandersypen2001Experimental,lu2007demonstration,lanyon2007experimental,politi2009shor,matthews2009compiled}. Initial implementations were based on nuclear magnetic resonance (NMR) and mostly followed Shor's original algorithm. For example,  15 was factorised using an NMR-based 7-qubit  quantum computer \cite{Vandersypen2001Experimental}. Interestingly, largest number that was factored using Shor's algorithm until 2012 was 21 and a 10 qubit quantum computer was used for the purpose \cite{lucero2012computing}. Due to the fact that considerably large quantum registers are required in Shor's original algorithm, now it's not usually used in its original form \cite{Vandersypen2001Experimental}. Though, Shor's algorithm, in principle, guarantees  factorization of a bi-prime number in polynomial time, the requirement of very large quantum registers restricted its applicability in factorising relatively large bi-primes. This fact motivated researchers to look for alternative approaches. One such approach is to use a hybrid scheme, variants of this approach are reported in Refs. \cite{peng2008quantum, pal2019hybrid, li2017high}. These variants are quite close to each other and each of them require relatively less quantum resources than that required in Shor's original approach. Here, it would be apt to note that hybrid schemes refer to those schemes, where part of the factorisation task is done classically to reduce the requirement of quantum resources which are costly. 

In 2008, Peng et al. \cite{peng2008quantum}, devised an algorithm utilizing adiabatic quantum computing, on the basis of the work of Farhi et al. \cite{farhi2001quantum} and demonstrated factorization of 21 using 3-qubits. Furthermore, in 2012, Xu et al., have improved the scheme by solving some equations mathematically. They have further demonstrated the beauty and power of the factorisation algorithms of this class by factorizing 143 using a 4-qubit NMR quantum processor\cite{xu2012quantum}. Two years later, Dattani and Bryans established that Xu et al., had actually factored 3599, 11663, and 56153, but could not recognize that. In the work of Dattani and Bryans, classical resources were used for partially simplifying a set of bit-wise factoring equations which allowed them to reduce the quantum overhead for a set of numbers \cite{dattani2014quantum}. In  2019, Pal et al.,  have demonstrated a hybrid  scheme for factorization of 551 using 3-qubit system  resources  \cite{pal2019hybrid}. 
The progress has been continuing and very recently (in 2017),
factorization of 35 was demonstrated using a single solid spin system under ambient conditions by Xu et al., \cite{xu2017experimental}. Furthermore, a set of relatively large numbers have recently been experimentally factorized. Specifically, combining the concepts of quantum annealing and computational algebraic geometry, a new approach for quantum factorization is developed by Dridi and Alghassi \cite{dridi2017prime} and the same has been successfully used to factorize all bi-primes up to 200099 using the D-Wave 2X processor \cite{anschuetz2019variational} and the experimental factorization of 291311 is performed by Li et al., \cite{li2017high} using a hybrid adiabatic quantum algorithm.

The importance of the factorization problem and the facts that (i) quantum factorization has yet been performed using only a few potential candidates for the scalable quantum computer and (ii) hybrid schemes have potential to factorize large bi-primes using small and noisy quantum computers available today,  have motivated us to modify the hybrid scheme given in \cite{pal2019hybrid,xu2012quantum} to obtain a new scheme which can be implemented in another experimental platform, namely IBM Quantum processor. Specifically, the algorithm proposed here is designed for factorization of bi-primes using a Josephson-qubit based quantum computer \cite{IBMQE,devitt2016performing}.
Interested readers may find a detailed user guide on how to use this computer at \cite{IBMQE},
and a lucid description of the working principle of a Josephson-qubit based quantum computer in Ref. \cite{steffen2011quantum}. This quantum computer was placed in the cloud in 2016. It immediately drew considerable attention of the quantum information processing community, and several quantum information tasks have already been performed using this quantum computer. Specifically, in the
domain of quantum communication, properties
of different quantum channels that can be used for quantum communication have been studied experimentally \cite{wei2018efficient} and experimental realizations of a
quantum analogue of a bank cheque \cite{behera2017experimental}
that is claimed to work in a banking system having a quantum network,
and solving set of linear equations \cite{doronin2020solving}
and two-qubit quantum states using optimal resources \cite{sisodia2017design}, have been reported; in the field of quantum foundation, violation
of multi-party Mermin inequality has been realized for 3, 4, and 5
parties \cite{alsina2016experimental}; an information
theoretic version of uncertainty and measurement reversibility has
also been implemented \cite{berta2016entropic}; in 
the area of quantum computation, a comparison of two architectures
using demonstration of an algorithm has been performed \cite{linke2017experimental},
a quantum permutation algorithm \cite{yalccinkaya2017optimization},
and a quantum eigensolver method based experimental search for ground state energy  energy for molecules of increasing  size  up  to   $\rm{BeH_2}$   \cite{kandala2017hardware} have been implemented recently. Further, a non-abelian braiding of surface code defects \cite{wootton2017demonstrating} and a compressed simulation of the transverse field one-dimensional Ising interaction (realized as a four-qubit Ising chain that utilizes only two qubits) \cite{hebenstreit2017compressed}
have also been demonstrated. Clearly, the IBM quantum computer
has already been used for the experimental realizations of various
quantum information processing tasks. However, to the best of our
knowledge, IBM quantum computer has not yet been used to realize Shor's algorithm or hybrid algorithms for factorization. This paper aims to address that gap.

This paper is organized as follows. 
Sec. \ref{intro} sets motivation behind choosing factorization problem, followed by detailing the gradual development of hybrid (i.e., combined classical and quantum) strategies to obtain efficient solution. In Sec. \ref{theory}, we revisit the general scheme for hybrid factorization. In Sec. \ref{fact35}, we elaborate on a specific case-  factorization of 35 using hybrid scheme of factorization, to be precise, we construct bit-wise equations and bit-wise matrix, optimize them to calculate unknown constants, and obtain relation between bit variables, formulate corresponding problem Hamiltonian using the procedure given in Sec. \ref{theory}. In Sec. \ref{exp}, we illustrate experimental implementation of the quantum part of the hybrid scheme for factorizing 35 using IBM architecture. For the purpose, we use Adiabatic evolution for ground state search of the problem Hamiltonian constructed in the previous section, which is the desired solution. In Sec. \ref{results}, we show the experimental results, which reveals the unknown qubit state and hence one factor of the composite number, and ultimately We conclude in Sec. \ref{concl}.

\section{Theory} \label{theory}

Let's consider a $b_{n}$-bit number $N=\sum_{j=0}^{b_{n}-1}2^{j}n_{j}$. Although, there may be various sets of factors of a composite number of bit length $b_{n}$ with maximum number of possibilities equals $\lceil \frac{b_{n}}{2} \rceil$. Here, $\lceil x \rceil$ corresponds to a  ceiling function. Acquainted with the requirement of cryptosystems, here we consider  $N$ as a distinct bi-prime. Let us assume that the two factors of $N$ are $P=\sum_{k=0}^{b_{p}-1}2^{k}p_{k}$ and $Q=\sum_{l=0}^{b_{q}-1}2^{l}q_{l}$ with bit length $b_{p}$ and $b_{q}$ respectively, such that, either $b_n=b_p+b_q$ or $b_{n}=b_{p}+b_{q}-1$. From the definition of prime factors, $p_{0}=q_{0}=1$  and $p_{b_{p}-1}$ = $q_{b_{q}-1}$=1. Thus, identity $N=PQ$ becomes,
\begin{equation}
\sum_{j=0}^{b_{n}-1}2^{j}n_{j} = \sum_{k=0}^{b_{p}-1}2^{k}p_{k} \sum_{l=0}^{b_{q}-1}2^{l}q_{l}. \label{eq1}
\end{equation}
As follows from the above equation either $b_{n}=b_{p}+b_{q}$ or $b_{n}= b_{p}+b_{q}-1$. The hybrid scheme for factorization can be divided into the following steps.
\begin{itemize}
\item Formulating multiplication table for $P$ and $Q$.
\item Constructing bit-wise equations from multiplication table.
\item Simplifying bit-wise equations using classical computation.
\item Constructing bit-wise Hamiltonian and hence problem Hamiltonian.
\item Obtaining unitaries corresponding to adiabatic evolution starting from given Hamiltonian to the problem Hamiltonian.
\item Decomposition of a given unitary using gates available in IBM Clifford library.
\item Obtaining solution of problem Hamiltonian using adiabatic quantum computation.
\end{itemize}

%
In the following, we illustrate the hybrid scheme for factorization with an example of $N=35$. We would also report experimentally obtained factors using this scheme. Experimental implementation of quantum part of the scheme, i.e., constrained minimization using adiabatic evolution has been done in 5-qubit IBM quantum processor involving superconducting qubits. 
%
%
%
\subsection{Simplification of bit-wise constraint equation}
For the purpose of bit-wise comparison of coefficients of both sides, we need to rewrite the above equation by introducing new index $c=k+l$. In terms of this new index $c$ modified equation becomes  
\begin{equation}
\sum_{j=0}^{b_{n}-1}2^{j}n_{j} = \sum_{c=0}^{b_{p}+b_{q}-2}2^{c} \sum_{l=c_{min}}^{c_{max}}p_{c-l}q_{l}.
\label{rewritten_m}
\end{equation}
Here, $c_{min}= max(0,c-lp-1)$ and $c_{max}= min(c,lq-1)$. Further, term $p_{c-l}q_{l}$ can be broken as $p_{c-l}q_{l}+C_{c-1,c}=s_{c,l}+2C_{c,c+1}$. Here, $C_{c-1,c}$ is the carry from $(c-1)^{th}$ column to $c^{th}$ column and $C_{c,c+1}$ is the carry from $(c)^{th}$ column to $(c+1)^{th}$ column. Such a decomposition allows us an easy understanding of the construction of the multiplication table. In order to get simplified bit-wise factoring equations, without any loss of generality, we now add all carries in the given column. So the bit-wise factoring equation takes following form
\begin{equation}
\sum_{l=c_{min}}^{c_{max}}p_{c-l}q_{l}+C_{c}-2C_{c+1}= n_{j}.
\label{bitequation}
\end{equation}
Here, cumulative carry $C_{c}= \sum_{c_{min}}^{c_{max}}C_{c-1,c}$ and $n_{j}$ is the bit value of number $N$ for the $j^{th}$ order (power) of the base value in the binary system. 
\subsection{Constraint optimization using classical resources} \label{classicaloptimization}

Consider the following equation,
\begin{equation}
    \sum_{j=0}^{b_{n}-1}2^{j}n_{j} = \sum_{c=0}^{b_p+b_q-2} 2^{c}n_{c}+2C_{c+1}-C_{c} 
\label{lastequation1}
\end{equation}
and equation
\begin{equation}
 \sum_{j=0}^{b_{n}-1}2^{j}n_{j}= \sum_{c=0}^{b_p+b_q-2} 2^{c}n_{c}+ C_{b_{p}+b_{q}-1}
 \label{lastequation2}
\end{equation}
obtained after putting the value of $$\sum_{l=c_{min}}^{c_{max}}p_{c-l}q_{l}$$ from Eq.(\ref{bitequation}) into Eq.(\ref{rewritten_m}). Writing Eq. (\ref{lastequation2}) in such a way allows us to calculate values of carry $C_{b_{p}+b_{q}-1}$. A direct comparison of the L$.$H $.$S$.$ $\mathrm{with}$ R$.$H$.$ S$.$ reveals the values of the carry $ C_{b_{p}+b_{q}-1} $. Also, as there is no incoming carry to the first column, we set $C_{0}=0$. Moreover, bit-wise equation for $c=0$ and $c=b_{p}+b_{q}-2$, i.e., $1+C_{0}=1-2C_{1}$ and $C_{b_{p}+b_{q}-2}+1=n_{b_{p}+b_{q}-2}+2C_{b_{p}+b_{q}-1}$ give $C_{1}$ and $C_{b_{p}+b_{q}-2}$. Next we obtain the following equality by rewriting the bit-wise equation. 
\begin{equation}
    \mathrm{max}\lfloor C_{c+1} \rfloor = \lfloor \mathrm{max}(\frac{1}{2}\sum_{l=C_{min}}^{C_{max}} p_{c-l}q_{l}+\frac{1}{2}C_{i})-\frac{n_{i}}{2}  \rfloor.
\end{equation}
This equality can be used to calculate the upper bounds over $C_{j+1}$, and this upper bound can be iteratively used to obtain upper bound on next $C$ value. The bit equation obtained under these  constrain, further allows us to simplify complete set of bit equations with minimum number of independent parameters.  
For the two cases, namely, Case A $\colon$ when $b_n=b_p+b_q$ and Case B $\colon$ when $b_n=b_p+b_q-1$ the values of the carry are $1$ and $0$ respectively.

\subsection{Construction of problem Hamiltonian} \label{probHam}
In 2008, Peng et al. developed a framework for solving factorization problem using quantum adiabatic algorithm. For the purpose, they formulated the factorization problem as a minimization problem by constructing a function $f(p,q)=(N-pq)$. The solution of this equation should reveal the values of classical variables $p$ and $q$. They  further suggested that any corresponding quantum approach to the minimization problem must involve finding the ground state of the Hamiltonian, which can be considered to be of the form $H=f(p,q)\ket{p,q}\bra{p,q}$, where $f(p,q)$ is the ground state eigenvalue and $\ket{p,q}$ is the corresponding product state of states $\ket{p}$ and $\ket{q}$. The problem Hamiltonian for the factorization problem takes the form $H=(N\operatorname{I_{2^{n}}}-PQ)^2$, where $\operatorname{P}=\sum_{i}2^iA_{i}$ $\operatorname{Q}=\sum_{i}2^iA_{i}$ and $\operatorname{A_{i}}= \frac{\operatorname{I}-\operatorname{\sigma_{iz}}}{2}$ with eigenvalues 0 and 1 for eigenstates $\ket{0}$ and $\ket{1}$, respectively. Furthermore, Xu et al. \cite{xu2012quantum} have used another approach to construct a Hamiltonian which uses relatively less quantum resources than that used in Ref. \cite{peng2008quantum} but still uses more resources than used by Pal et al. in Ref. \cite{pal2019hybrid}. In this article, in order to construct the problem Hamiltonian we have used the same approach as was used by Xu et al. in Ref. \cite{xu2012quantum} i.e., to begin with we have transformed the classical bit variable $p_{i}$ and $q_{i}$ into operators such that $p_{i} \rightarrow A_{i}$ and $q_{i} \rightarrow A_{i+b_{k}-2}$.  

 \subsection{Quantum adiabatic algorithm for ground state search}
Quantum adiabatic algorithm states that, during the evolution of a quantum system under a slowly varying (as per adiabaticity condition \cite{farhi2000quantum}) time dependent Hamiltonian $H(t)$, system remains in the same eigenstate in which the system is prepared initially \cite{mesiah1961quantum}. Given a problem, adiabatic quantum computation typically involves encoding the solution to the problem in the ground-state of the  final Hamiltonian. A suitable initial Hamiltonian $H_{i}(0)$ is chosen for which ground-state can be prepared easily and evolved to the final Hamiltonian $H_{f}(T)$. 
Then the Hamiltonian of the system is slowly varied such that the system stays in the ground state of the instantaneous Hamiltonian. The instantaneous Hamiltonian can be designed as an interpolation (linear or nonlinear) between the initial Hamiltonian and the final Hamiltonian \cite{farhi2000quantum}. For the linear interpolation parameter $s=\frac{t}{T}$, such that $0 \leq s \leq 1$, where $t$ is the evolution time and $T= \vert \frac{max(\frac{dH(s)}{ds})}{\epsilon \Delta^2/\hbar} \vert$, is the total evolution time. The adiabatic theorem guarantees that system will evolve to the ground state of the final Hamiltonian with probability $1-\epsilon^2$, and the transformed Hamiltonian would become
\begin{equation}
\operatorname{H(s)} = (1-s)\operatorname{H_i} + s\operatorname{H_f}. 
\end{equation}
Considering the Hamiltonian as piecewise constant Hamiltonian with M pieces the time ($t$) can be rewritten as $t=\frac{m}{M}T$, where $0 \le m \le M$ and the Hamiltonian for the $m^{th}$ piece is
\begin{equation}
H_{m} = (1-\frac{m}{M})H_{i} + (\frac{m}{M})H_{f}.
\end{equation}
The unitary evolution $U_{m} = \exp{(-\iota H_{m}\delta t)}$ governed by the corresponding Hamiltonian $H_{m}$, where $t$ is the duration of $m^{th}$ piece of evolution. Unitary operation for the total evolution is $U=\prod_{m=1}^{M} U_{m}$.  

\section{Factorization of 35 using IBM's 5-qubit quantum processor} \label{fact35}
In order to demonstrate the method, we take the example of 35. As mentioned in Sec. \ref{theory} there are two possible cases for the choice of the bit-length of the bi-prime factors, we start with the case $b_n=b_p+b_q$, thus $b_{p}=3$ and $b_{q}=3$. Although, one can start with any of the two cases and in case of obtaining an inconsistent solution in the first case, consistent solution will be guaranteed in the other case. We start by obtaining the multiplication table (see Tab. \ref{bitwiseMultiplication})  for the composite number $N=35$. 

\begin{table}[b]
\begin{tabular} { | c |c| c| c| c | c | c | c | c | c | c | c | c |} 
\hline
 \rm{ \textbf {j}} $\rightarrow$&5&4&3&2&1&0\\
\rm{\textbf{l}} $\downarrow$&&&&&&\\
\hline
&&&&1&$p_{1}$&1\\

&&&&1&$q_{1}$&1\\
\hline
0&&&&1 & $p_{1}$&1\\
\hline
1&&& $q_{1}$&$p_{1}q_{1}$&$q_{1}$&\\
\hline
2&&$1$ & $p_{1}$ &  $1$&&\\
\hline
Carry&$c_{4,5}$ & $c_{3,4}$&$c_{2,3}$&$c_{1,2}$&&\\
& & $c_{2,4}$&&&&\\
\hline
Cumulative Carry &&&&&&\\
$C_{c}$&$C_{5}$&$C_{4}$&$C_{3}$&$C_{2}$&$C_{1}$&$C_{0}$\\
\hline
$n_{j}$&1&0&0&0&1&1\\

\hline
\end{tabular}
\caption{Bit-wise multiplication table for 35. Columns in the table correspond to parameter $c$ introduced to get simplified bit-wise equations (see Eq. (\ref{lastequation2})) while rows correspond to $l$ values (see Eq. (\ref{lastequation1})). Bit values $n_{j}$ for $j=0\colon5$ are provided in the last row of the table.}
\label{bitwiseMultiplication}
\end{table}

The bit-wise equations obtained from the  multiplication table (Table \ref{bitwiseMultiplication}) are$\colon$
\begin{align*}
 1+C_{0}=1,\\
    p_{1}+q_{1}+C_{1}-1=2C_{2},\\
    1+p_{1}q_{1}+1+C_{2}-0=2C_{3},\\
    p_{1}+q_{1}+C_{3}-0=2C_{4},\\
    1+C_{4}-0=2C_{5},\\
    C_{5}=1.\\    
\end{align*}
We then optimize above set of equations using the constrain optimum condition given in Sec. \ref{classicaloptimization}. The carries thus obtained are $C_{0}=0, C_{1}=0, C_{2}=0,C_{3}=1, C_{4}=1, C_{5}=1$. This leaves us with only one bit equation, i.e., $p_1+q_{1}=1$. We then construct the operators corresponding to bit values $p_{1}$ and $q_{1}$ as discussed in \ref{probHam}. Thus, the operators corresponding to the bit values $p_{1}$ and $q_{1}$ are $P=Q=\sum_{i}A_{i}=A_{1}$ and for $A_{1}= \frac{\operatorname{I}-\operatorname{\sigma_{1z}}}{2}$. Now the problem Hamiltonian becomes 

\begin{align*}
H_{p}= (P_{1}+Q_{1}-1)^2 \\       
     = (\frac{\operatorname{I}-\operatorname{\sigma_{z}}}{2}+\frac{\operatorname{I}-\operatorname{\sigma_{z}}}{2}-\operatorname{I})^2\\
     = {\operatorname{\sigma_{z}}}^2\\
     = {\operatorname{I}}.    
\end{align*}

We now use quantum adiabatic evolution for finding the ground state of the final Hamiltonian $H_{p}$ starting from the ground state of the easily initializable Hamiltonian in the IBM's $QX_{4}$ processor, i.e., $H_{i}=J \sigma_{z}$, in the units of $\hbar$ and $J \approx 2\pi\times10^{6}$ rad/sec. The ground state of the initial Hamiltonian $H_{i}$ is $\ket{-}=\frac{\ket{0}-\ket{1}}{2}$ and ground state of the final Hamiltonian $H_{f}$ is degenerate and are $\ket{0}$ and $\ket{1}$ with eigenvalues 1. We decided to adiabatically evolve the Hamiltonian in 8 steps. During each step the Hamiltonian can be considered as piecewise constant and the corresponding unitary operators can be obtained using the formula $U_{m}=\exp{(-\iota H_{m} \Delta t)}$, where $H_{m}$ is the Hamiltonian in the $m^{th}$ step and $\Delta t$ is the duration of the step. 

\begin{figure}[h]
\includegraphics[width=8cm,trim=5cm 3cm 7cm 3cm]{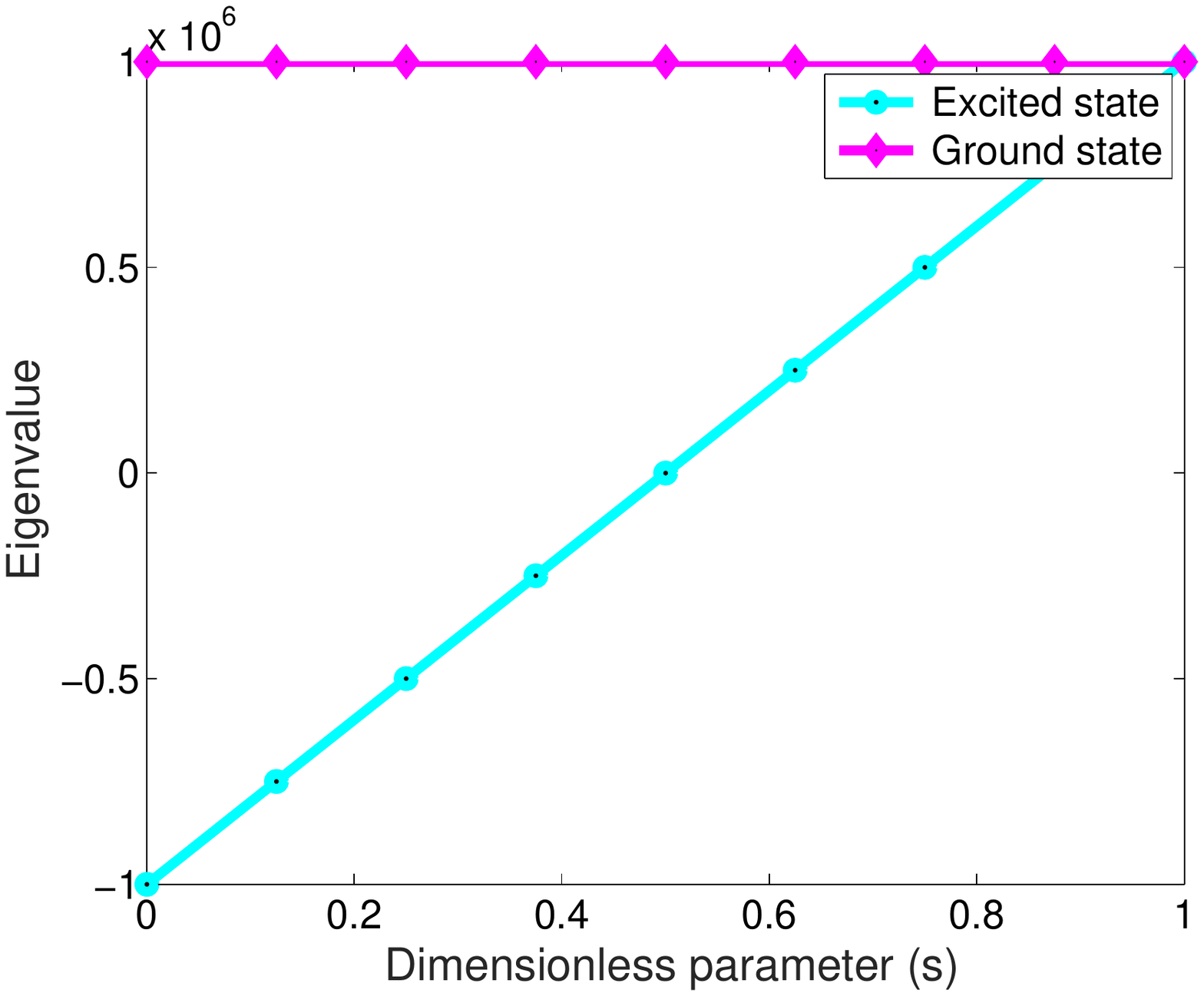}
\caption{\label{fig:Adb}(Color online) Simulation shows no anticrossing between two states of the qubit q[0] of the IBM's $QX_{4}$ quantum processor during adiabatic evolution for time chosen $T=10 \mu s$ in 8 steps. The Hamiltonian used are $H_{i}= J \sigma_{z}$ and $H_{f}= J\operatorname{I}$.}
\end{figure}

We first optimize the number of steps to check the adiabaticity condition being satisfied by the given set of Hamiltonians i.e., we check if the ground state of the initial Hamiltonian reaches the ground state of the final Hamiltonian without anticrossing as  shown in the Fig. \ref{fig:Adb}. 

\subsection{Decomposition of Unitaries}
We now decompose the unitaries obtained in Sec. \ref{fact35} for each of the 8 steps into the single-qubit Clifford+T gates. In actual implementation of Adiabatic evolution of the system from the ground state of the initial Hamiltonian to the ground state of the final Hamiltonian. The actual decomposition for a general unitary is shown in Eq. (\ref{nine}) and the exact values of $\theta$s for each step are given in Tab. \ref{theta}.  In this stage, we need to be specific to the quantum processor to be used, as the available gate library and the ease with which different gates can be realized in a particular implementation/architecture are different. Here, we are interested in implementing  the proposed scheme using an IBM QX4 processor, which restricts us to decompose the unitaries in terms of the available Clifford gate library of IBM Quantum Experience (QE). To be specific, to implement our scheme in IBM's QX4, any general unitary which we require to implement as part of our scheme (circuit), has to be decomposed in terms of the Clifford+T gates. In general, a single-qubit unitary obtained with the chosen initial and obtained final Hamiltonians are of the form, 
$U = \left(
\begin{array}{cc}
    a & 0 \\
    0 & b
\end{array}\right)$.
Each element in the given unitary is a complex number. Thus, $ a= r_{1}\exp(\iota \theta_{1}) $ and,  $ b= r_{2}\exp(\iota \theta_{2})$ hence corresponding to each unitary we have two matrices corresponding to r and $\theta$ values i.e., $R = \left(  \begin{array}{cc}r_{1} & 0 \\ 0 & r_{2} \end{array} \right)$ and $ \Theta = \left(  \begin{array}{cc} \exp(\iota \theta_{1}) & 0 \\ 0 & \exp(\iota \theta_{2}) \end{array} \right)$ such that 
\begin{equation}
\operatorname{U}= \operatorname{R} \cdot \operatorname{\Theta}.
\end{equation}
The values of $\theta$ are given in Tab. \ref{theta}. For the case of 35, $R$ matrices for all unitaries are identity, so $\operatorname{U}= \operatorname{\Theta}$.
The IBM Clifford+T gate library consists of following single-qubit unitaries:\\
(i) The Pauli gates: $\operatorname{I}$, $\operatorname{X}$, $\operatorname{Y}$, and $\operatorname{Z}$\\
(ii) General gates: $\operatorname{U_{1}\left(\theta \right)}$, $\operatorname{U_{2}}(\phi,\lambda)$, and $\operatorname{U_{3}}(\theta,\phi,\lambda)$.\\
(iii) Phase Gates: $\operatorname{S} (\operatorname{S^{\dagger}})$, $\operatorname{T} (\operatorname{T^{\dagger}})$, and\\
(iv) Other Gates: $\operatorname{H}$ (Hadamard) \\
In what follows, we will use phase gate and Pauli $X$-gate to construct unitaries to be implemented in our experiments. The decomposition of $\Theta$ in Clifford+T gate library can be obtained as

\begin{widetext}
\begin{eqnarray}
\operatorname{\Theta} &=&\left({\begin{array}{cc} 1 & 0\\ 0 & \exp(\iota \theta_{2})\end{array}}\right)  \left({\begin{array}{cc}  0 & 1\\ 1 & 0\end{array}}\right)\left({\begin{array}{cc} 1 & 0\\ 0 & \exp(\iota \theta_{1})\end{array}}\right) \left({\begin{array}{cc}  0 & 1\\ 1 & 0\end{array}}\right)\\   \nonumber
&=& \operatorname{U_{1}}\left(\theta_{2}\right) \cdot \operatorname{X} \cdot \operatorname{U_{1}}\left(\theta_{1}\right) \cdot \operatorname{X}.
\label{nine}
\end{eqnarray}
\end{widetext}
Combining $R$ and $\Theta$ matrices, the unitary U$_{m}$ for m$^{th}$ step can be written as
 \begin{equation}
U_{m} = \operatorname{U_{1}^{m}}\left(\theta_{1}^{m}\right) \cdot \operatorname{X} \cdot \operatorname{U_{1}^{m}}\left(\theta_{2}^{m}\right) \cdot \operatorname{X}, 
\label{um}
 \end{equation}
where, $\theta_{1}^{m}$ and $\theta_{2}^{m}$ are the angles in the m$^{th}$ unitary.
Thus, the total unitary for quantum part of the hybrid factorization scheme is $\operatorname{U}= \prod_{m=8}^{m=1} U_{m}$.
Here, it is important to mention that gates in the unitary have been applied in the right to left order. The values for $\theta_{1}^{m}$ and $\theta_{2}^{m}$ for different steps are provided in Tab. \ref{theta}.

\begin{center}
\begin{table}[b]
\begin{tabular}{|m{.5cm}|m{.8cm}|m{.8cm}|m{1.5cm}|m{1.5cm}|}
\hline
m & $r_{1}$ & $r_{2}$ & $\theta_{1}^{m}$ & $\theta_{2}^{m}$\\
\hline
1 & 1 & 1 & -1.2500 & 0.9375\\
\hline
2 &  & 1 & -1.2500 & 0.6250 \\
\hline
3 & 1 & 1 &-1.2500 & 0.3125\\
\hline
4 & 1 & 1 & -1.2500 & 0 \\
 \hline
5 & 1 & 1 & -1.2500 & -0.3125\\
 \hline
 6& 1 & 1 & -1.2500 & -0.6250\\
 \hline
 7 & 1 & 1 & -1.2500 & -0.9375 \\
 \hline
 8 & 1 & 1 & -1.2500 & -1.2500\\
  \hline
\end{tabular}
\caption{The r and $\theta$ values for unitaries in each step of adiabatic evolution.} \label{theta}
\end{table}
\end{center}

\begin{figure}[h]
\includegraphics[width=5cm,trim=9cm 3cm 10cm 1cm]{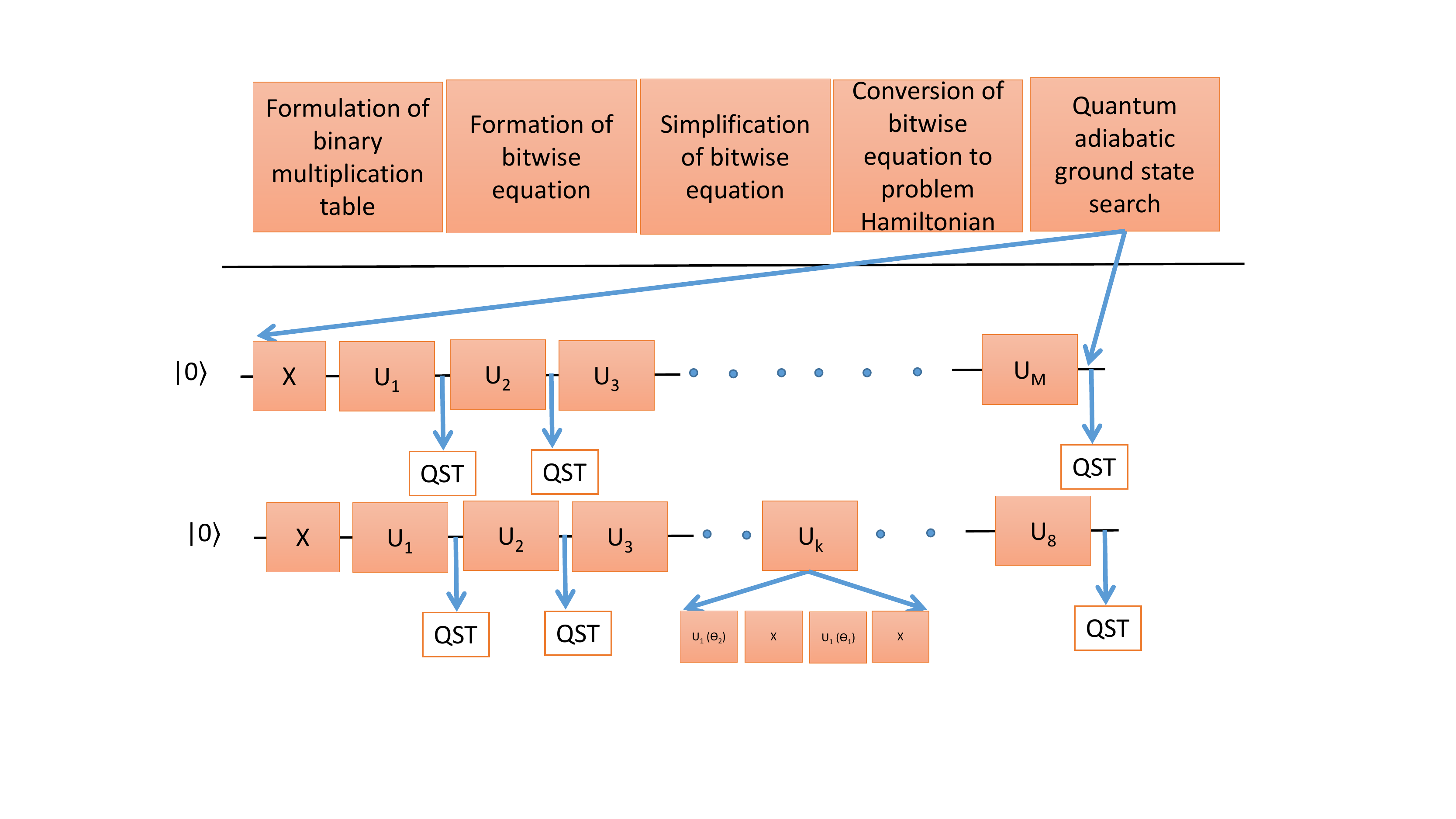}
\caption{(Color online) Schematic procedure of complete scheme for hybrid factorization. Top trace shows the steps of the scheme. Second trace shows circuit for obtaining ground state of the final Hamiltonian $H_{f}$ starting from $H_{i}$ through adiabatic evolution. The trace below that shows quantum circuit for implementing quantum adiabatic evolution part on IBM's QX4 processor for factorization of 35, by initializing the qubit system to the ground state i.e., $\ket{1}$ of the initial Hamiltonian $J \sigma_{z}$ and the lowest trace shows the gate decomposition in IBM's gate library of $k^{th}$ unitary $U_{k}$. }
\label{schematic}
\end{figure}

\section{Experiments} \label{exp}
 This experiment has been performed on an open access 5-qubit quantum processor placed on cloud by IBM corporation. In particular, we have used the architecture of IBM's QX4 (IBM Q 5 Tenerife)\cite{IBMQE}, which consists of superconducting Transmon qubits \cite{malkoc2013quantum}. A schematic diagram of this architecture and description of the architecture can be found in \cite{sisodia2017design,sisodia2017experimental} and references therein. The basic gate library used for the single-qubit gates are $\operatorname{H}$, Pauli operators $\operatorname{X}$, $\operatorname{Y}$, $\operatorname{Z}$, parametric gates $\operatorname{U_{1}}$, $\operatorname{U_{2}}$, and $\operatorname{U_{3}}$. The operator $\operatorname{U_{1}}$ depends on single parameter $\theta$, operator $\operatorname{U_{2}}$ depends on two parameters $\theta, \phi$, and operator $\operatorname{U_{3}}$ depends on three parameters $\theta, \phi, \lambda$. We chose qubit q[0] for implementation of quantum adiabatic evolution of ground state search. We initialize the system to the ground state of the initial Hamiltonian by applying the $\operatorname{X}$ gate ($\sigma_{x}$) to the qubit q[0]. To evolve this state adiabatically we apply a set of eight unitaries, in their decomposed form as shown in the previous section. In order to extract the probabilities of the final state we perform quantum state tomography after each step. The directly measured observable in IBM processor are $\ket{0}\bra{0}$ and $\ket{1}\bra{1}$ which allow us to calculate $\expv{Z}$. This is sufficient to reveal the probabilities $p_{0}$ and $p_{1}$. In order to measure real and imaginary part of the coherence term, we have used the method described in our earlier work \cite{shukla2020complete} i.e., we have applied, $H$ gate followed by $\operatorname{Z}$-measurement to reveal real part and $S^{\dagger}H$ followed by $\operatorname{Z}$-measurement for measuring its imaginary part.

\section{Results} \label{results}
\begin{figure}[h]
\includegraphics[width=7cm,trim=5cm 9cm 5cm 8cm]{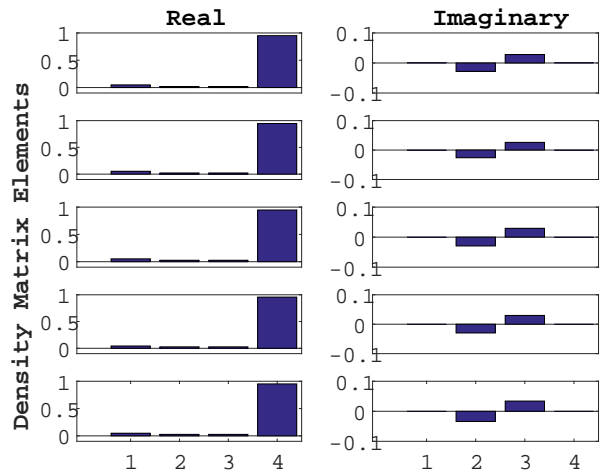}
\caption{\label{fig:results1} (Color online)  Figure shows density matrices in rows, with the columns representing the real and imaginary parts of the density matrix. Top row indicates the initial state, subsequent rows represent state at even instances of applying unitaries i.e., 2, 4, 6, and 8 times, while going down.   Ticks 1, 2, 3, 4 correspond to the elements $\ket{0}\bra{0}$, $\ket{0}\bra{1}$, $\ket{1}\bra{0}$, $\ket{1}\bra{1}$.}
\end{figure}

\begin{figure}[h]
\includegraphics[width=9cm,trim=3cm 8cm 3.5cm 8cm]{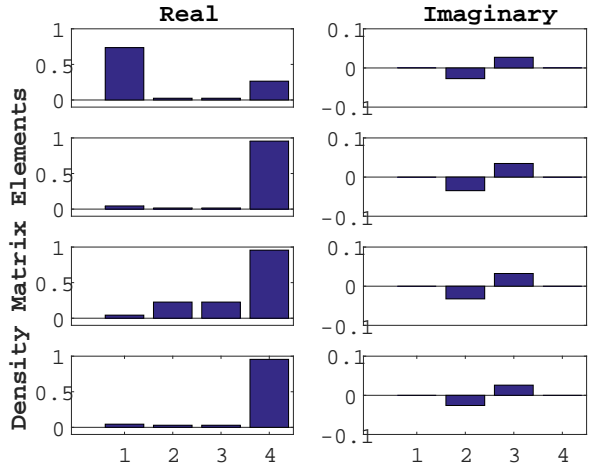}
\caption{\label{fig:results2} (Color online) Figure shows density matrices in rows, with the columns representing the real and imaginary parts of the density matrix. Rows represent state at odd instances of applying unitaries to the initial state i.e., 1, 3, 5, and 7 times, while going down. Ticks 1, 2, 3, 4 correspond to the elements $\ket{0}\bra{0}$, $\ket{0}\bra{1}$, $\ket{1}\bra{0}$, $\ket{1}\bra{1}$. The result reveals the system to be in the ground state of the final Hamiltonian.}
\end{figure}

The QST results, are shown in Fig. \ref{fig:results1} and Fig. \ref{fig:results2}. To measure the final state we have the corresponding measurement operator in the Clifford library. The ground state of the final Hamiltonian provides us the solution to our problem, in our case, we would obtain probability of  $p_{0}$ and $p_{1}$ for the states $\ket{0}$ and $\ket{1}$, respectively, after performing the experiment 8192 times (i.e., the maximum number of runs that one can select from the interface provided by IBM QE). The tomography results reveal that after the full adiabatic evolution of the system in 8 steps, system is in the ground state of the final Hamiltonian. Since the Hamiltonian at this point is degenerate, consequently, solution of the problem Hamiltonian are any of the two states $\ket{0}$ and $\ket{1}$. If we consider $\ket{0}$ as the solution, then the corresponding classical bit value would be $p_{1}=0$ leading to the first factor of the composite number 35 in the binary system as $1 p_{1} 1 =101$ and consequently, in decimal system as $P=\sum_{k} 2^{k} p_{k}=5$. The conjugate bit value by using the identity $p_{1}+q_{1}=1$ is $q_{1}=1$ and the corresponding prime factor in the binary system is $1 q_{1} 1=111$, and consequently the number, in the decimal system would be $Q=\sum_{l} 2^{l} q_{l}= 7$. The two factors can also be obtained in the same way if we consider $\ket{1}$ as the solution of the ground state search of the adiabatic evolution. 

\section{Conclusions} \label{concl}
As discussed above, factorisation of bi-prime numbers are important for hacking of RSA type cryptographic schemes and various related problems. However, factorisation of some bi-primes are not as difficult as that of the others. Specifically, for an even bi-prime, we already know that one factor is 2 and it's trivial to find the other factor. Further, there are excellent algorithms for finding square root, so factorisation of square bi-primes are easy. What we are left with is odd and square-free bi-primes which are used in cryptography. Here, we report a quantum-classical hybrid scheme for factorization of such (i.e., odd and square-free) bi-prime numbers.  The scheme proposed here is hybrid in nature, implying that the scheme utilizes both classical optimization techniques and adiabatic quantum optimization techniques. The advantage of such hybrid schemes underlies in the fact that they require less quantum resources (which are fragile and costly at the moment) in comparison with the purely quantum schemes designed for the same purpose. For, example, it's already understood that the extremely large quantum registers required in Shor's original protocol are not required in the hybrid schemes proposed later on. The same is true for our scheme as well as the schemes \cite{pal2019hybrid,xu2012quantum}  which have been extended here. Thus, in short the proposed scheme has the capability to factorise relatively large odd and square-free bi-primes using a small amount of quantum resources. To illustrate this, we have explicitly performed factorisation of 35 (which is an odd and square free bi-prime) using the smallest quantum computer available on the cloud (i.e., a five qubit quantum processor called IBM's QX4).  The quantum processor used here is known to be noisy, but here we have correctly obtained the prime factors of 35  with some small signatures of noise depicting the strength of the algorithm. We conclude the paper with a hope that with the availability of larger quantum processors, larger bi-primes will be factored using this algorithm and it will be found useful in the future development of the hybrid algorithm designing. 

\section{Acknowledgements}
Authors thank to Defense Research And Development Organization (DRDO), India for the support provided through the project number ERIP/ER/1403163/M/01/1603. Abhishek Shukla thanks to Applied Physics department, The Hebrew University of Jerusalem for the support.
\bibliographystyle{apsrev4-1}
\bibliography{ref}

\end{document}